# Comparative Analysis of Congestion Control Algorithms Using ns-2

Sanjeev Patel[1], P. K. Gupta[2], Arjun Garg[3], Prateek Mehrotra[4] and Manish Chhabra[5]

[1] Deptt. of Computer Sc. & Engg, Jaypee Institute of Information Technology,
Noida, UttarPradesh, 201307, India

[2] Deptt. of Computer Sc. & Engg, Jaypee University of Information Technology,
Waknaghat, Solan, Himachal Pradesh, 173215, India

[3] Deptt. of Computer Sc. & Engg, Jaypee University of Information Technology,
Waknaghat, Solan, Himachal Pradesh, 173215, India

[4] Deptt. of Electronics & Communication Engg., Jaypee University of Information Technology,
Waknaghat, Solan, Himachal Pradesh, 173215, India

[4] Deptt. of Electronics & Communication Engg., Thapar University,
Patiala, Punjab, 147004, India

**Abstract**
In order to curtail the escalating packet loss rates caused by an exponential increase in network traffic, active queue management techniques such as Random Early Detection (RED) have come into picture. Flow Random Early Drop (FRED) keeps state based on instantaneous queue occupancy of a given flow. FRED protects fragile flows by deterministically accepting flows from low bandwidth connections and fixes several shortcomings of RED by computing queue length during both arrival and departure of the packet. Stochastic Fair Queuing (SFQ) ensures fair access to network resources and prevents a busty flow from consuming more than its fair share. In case of (Random Exponential Marking) REM, the key idea is to decouple congestion measure from performance measure (loss, queue length or delay). Stabilized RED (SRED) is another approach of detecting nonresponsive flows. In this paper, we have shown a comparative analysis of throughput, delay and queue length for the various congestion control algorithms RED, SFQ and REM. We also included the comparative analysis of loss rate having different bandwidth for these algorithms.
**Keywords:** *Stochastic Fair Queing (SFQ), Random Early Detection (RED), Random Exponential Marking (REM), First In First Out (FIFO), Throughput, Delay, Queue length, Loss rate, and Utilization.*

## 1. Introduction

SFQ (Stochastic Fair Queuing) is a class of queue scheduling disciplines that are designed to allocate a pretty large number of separate FIFO queues [1]. Increasing the number of queues to a large extent helps to achieve fairness. RED queue management aims at alleviating this problem by detecting incipient congestion in advance and communicating the same to the end-hosts, allowing them to trim down their transmission rates before queues begin to overflow and packets start dropping. For this, RED maintains an exponentially weighted moving average of the queue length which it used as a congestion detection mechanism [2]. In order to be efficient, RED must ensure that congestion notification is conveyed at a rate which sufficiently suppresses the transmitting sources without underutilizing the link. RED must also ensure that the queue is configured with enough buffer space to hold an applied load greater than the link capacity from the time when congestion detection occurs to the time when the applied load reduces at the bottleneck link in response to the notification regarding congestion. FRED proposes to transform RED mechanisms to provide fairness by using per-active-flow accounting to make different dropping decisions for connections with different bandwidth usages [3]. When a flow persistently occupies a considerable amount of the queue's buffer space, it is identified and restrained to a smaller buffer space. Severity of congestion is indicated by queue lengths in various queue management algorithms. This inherent problem can be dealt by a fundamentally different active queue management algorithm, called BLUE [4]. BLUE has been shown to perform significantly better than RED both in terms of packet loss rates and buffer size requirements in the network. If buffer overflow causes the queue to recurrently drop packets, BLUE increments the marking probability, thus augmenting the rate at which congestion notification is sent back [4].

REM is an active queue management scheme that aims to achieve both high utilization and negligible loss and delay in a simple and scalable manner [5]. While congestion measure indicates excess demand for bandwidth and must track the number of users, performance measure, independently of the number of users, should be stabilized around their targets. The first idea of REM [5] attempts to match user rates to network capacity while clearing buffers, irrespective of number of users. The second idea embeds the sum of link prices (congestion measures), summed over all the routers in the path of the user to the end-to-end marking (or dropping) probability [5]. Number of active flows shares a linear relationship with number of different flows in the buffer. We simulated the network configuration having higher delay and lower bandwidth at the main bottleneck link [6]. In this paper, we used ns-2 network simulator. The structure of ns-2 and performance metrics considered in this paper has been given in section 2. In section 3, we described about network configuration and network parameters used in the simulation. This section also deals with implementation and result analysis observed in this simulation. Last section concludes with discussion of future work.

## 2. Performance Metrics

The complete NS class hierarchy has been shown in Figure 1. Now in the queue object of the hierarchy there are only two active queue management algorithm, RED and Drop tail. Now the other algorithm discussed in the section of network congestion control are implement under this object only. Queue serves as the parent of all these algorithms, and they are appended as a child to this element in the hierarchy. Similarly other can also be deployed in ns2 architecture and make them run [7].

### 2.1. Analysig Trace File

When the ns is run, the trace of each event can be stored in a trace file. While tracing into an output ASCII file, the trace is organized in 12 fields as shown in the following figure. Class hierarchy of ns2 and the description of each field is shown by their name and an sample example of trace file as given below in Figure 1 and Figure 2.

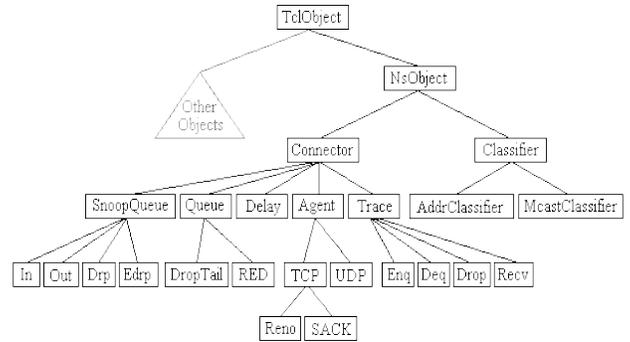

Fig. 1 Class Hierarchy

Fig. 2 Trace files structure

### 2.2 Packet Loss

Packets can be lost in a network because they may be dropped when a queue in the network node overflows. The amount of packet loss during the steady state is another important property of a congestion control scheme. The larger the value of packet loss, the more difficult it is for transport-layer protocols to maintain high bandwidths, the sensitivity to loss of individual packets, as well as to frequency and patterns of loss among longer packet sequences is strongly dependent on the application itself. This characteristic can be specified in a number of different ways, including loss rate, loss patterns, loss free seconds, and conditional loss probability. In this paper, we considered that packet loss would occur only due to the dropping of the packets. There is no loss due to other means.

### 2.3 Throughput

This is the main performance measure characteristic, and most widely used. This measure how soon the receiver is able to get a certain amount of data send by the sender. It is determined as the ratio of the total data received to the end to end delay. Throughput is an important factor which directly impacts the network performance.

2.4 Delay

Delay is the time elapsed while a packet travels from one point (e.g., source premise or network ingress) to another (e.g., destination premise or network degrees). The larger the value of delay, the more difficult it is for transport layer protocols to maintain high bandwidths. This characteristic can be specified in a number of different ways, including average delay, variance of delay (jitter), and delay bound. In this paper, we calculated end to end delay

2.5 Queue Length

A queueing system in networks can be described as packets arriving for service, waiting for service if it is not immediate, and if having waited for service, leaving the system after being served. Thus queue length is very important characteristic to determine that how well the active queue management of the congestion control algorithm has been working.

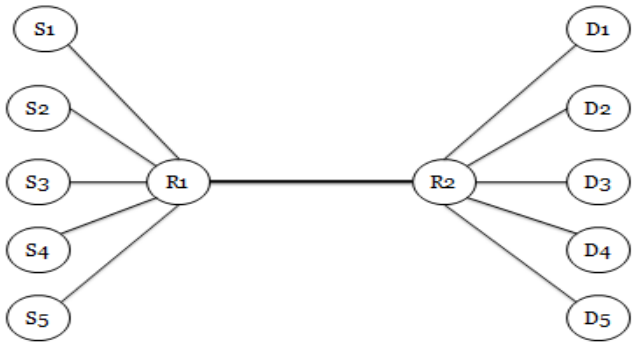

Fig. 3 Simulation Scenario

## 3. Simulations And Results

In this section, we discussed about network configuration used over the network simulator ns2 to simulate the three algorithms RED, SFQ and REM and after that we analyzed about the results obtained from our simulations. The algorithms compared here are first deployed into the ns2 architecture then following simulation scenario has been generated to compare their performance on the simulation setting as shown in Figure3.

3.1 Simulation Scenario

There are five nodes at each side of the bottleneck link. Here five nodes are acting as a TCP source and five nodes are acting as a TCP sink so that both routers are applying the congestion control algorithm. There is two- way traffic in the system. We consider the network scenario as shown in Figure 3. We simulate this network on ns2 for different AQM algorithms RED, SFQ and REM for same network parameters as given in Table 1 except to the bottleneck link. We simulated these three algorithms RED, SFQ, and REM on the same bottleneck link $R_1R_2$. Firstly we consider the bottleneck link to 5Mbps for each considered AQM algorithm. We considered a fixed packet size of 2 KB and buffer capacity of 4KB throughout the simulation. Round trip delay for each link has been displayed in Table 1. So it could be concluded from the Table 1 that minimum end to end delay should be larger than 60 ms. Our simulation has been observed over the period of 100 seconds. Whole simulation has been observed over small buffer capacity of 4KB.

3.2 Analysis of Loss Rate

Figure 4 shows about the loss rate occurred in RED, SFQ, and REM respectively. In our simulation, we vary the bandwidth of the bottleneck link as given in Figure 4 for each algorithm RED, SFQ, and REM. It has been observed that loss rate smoothly decreased as we are increasing the bandwidth of bottleneck link in case of RED. We got the drastic change in loss rate at 15 Mbps in case of SFQ because of unfairness achieved at this bandwidth. It has been concluded that SFQ and REM could achieved higher loss rate at higher bandwidth at some specific bandwidth but it could not be happen. It has been reflected more in case of SFQ. But RED shows smooth decrease in loss rate over increase in bandwidth.

3.3 Analysis of Throughput

It has been observed that REM had a best throughput and RED had least throughput among all these three algorithms for the simulation achieved at 5 Mbps of bandwidth. Figure 5 show that REM gets the good result and RED gets the poor result. It could be observed one point on throughput graph whenever smooth growth in throughput has been broken. It indicated about a starting point when dropping of packet took place. This achieved point in each algorithm has a same ratio as compared to their maximum achieved throughput.

Table 1:.Parameters for simulation

| Link | RTT (ms) | Rate (Mbps) | Protocol |
|---|---|---|---|
| S1 R1 | 10 | 100 | Drop tail |
| S2 R1 | 10 | 100 | Drop tail |
| S3 R1 | 10 | 100 | Drop tail |
| S4 R1 | 10 | 100 | Drop tail |
| S5 R1 | 10 | 100 | Drop tail |
| R1R2 | 40 | 10 | RED / SFQ /REM |
| R2D1 | 10 | 100 | Drop tail |
| R2D2 | 10 | 100 | Drop tail |
| R2D3 | 10 | 100 | Drop tail |
| R2D4 | 10 | 100 | Drop tail |
| R2D5 | 10 | 100 | Drop tail |

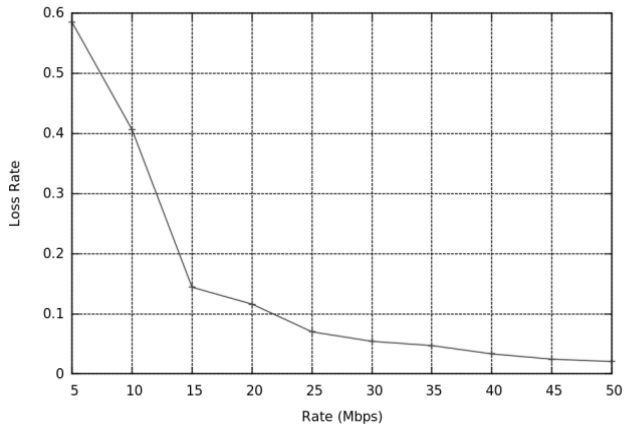

(a) RED

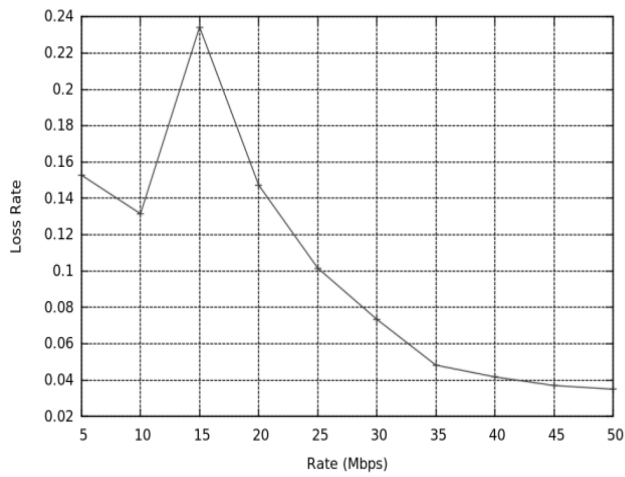

(b) SFQ

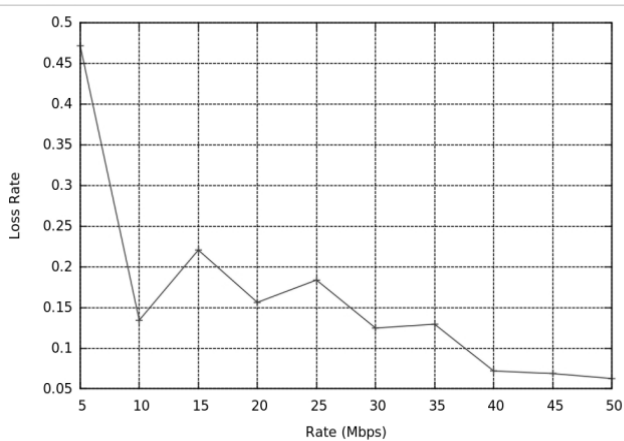

(c)REM

Fig. 4 Loss Rate for various algorithms (a) RED, (b) SFQ, (c) REM

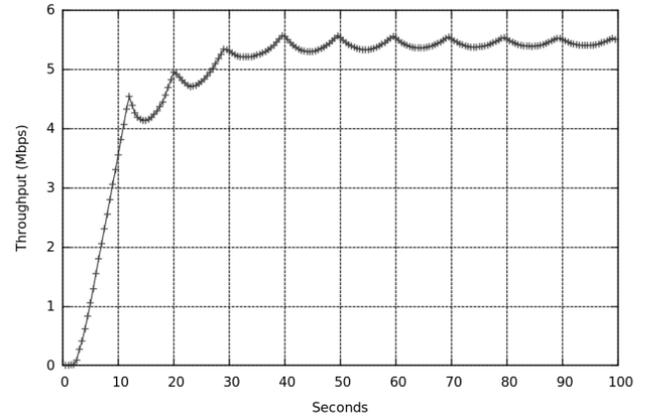

(a)RED

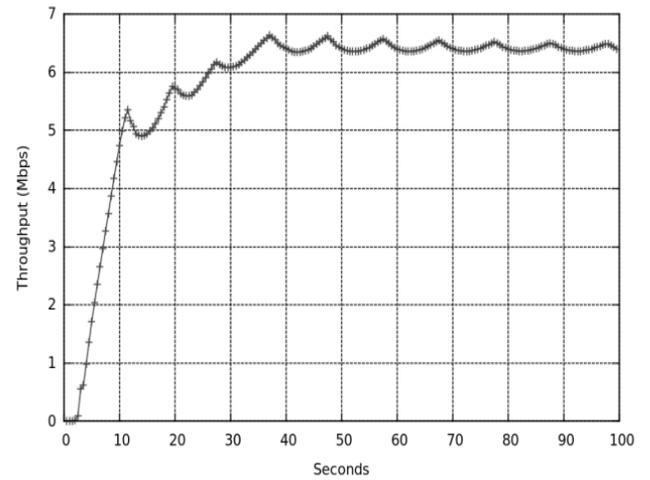

(b)SFQ

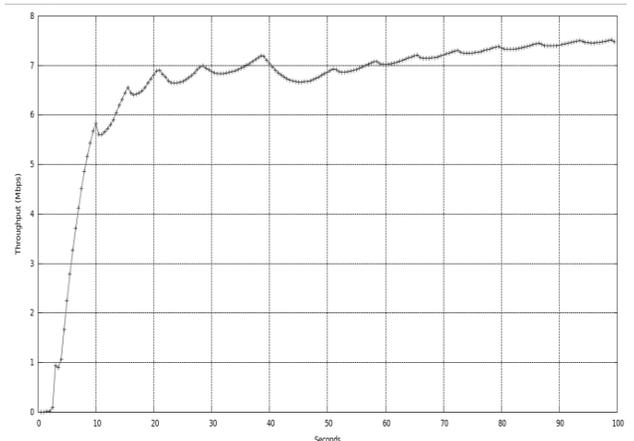

(c) REM

Fig. 5 Throughput Diagram for various algorithms (a) RED, (b) SFQ, (c) REM

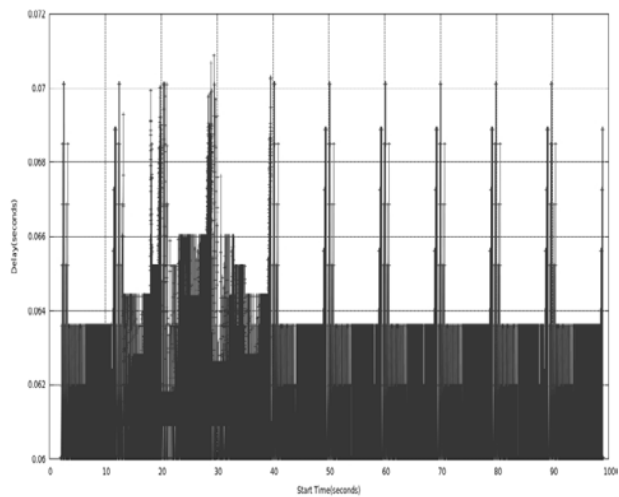

(a) RED

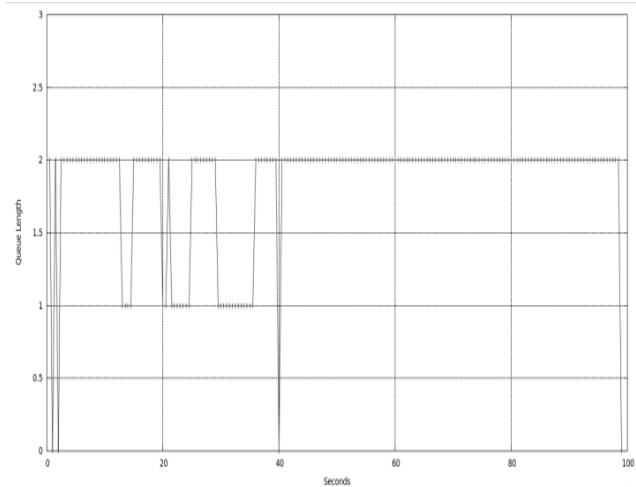

(a)RED

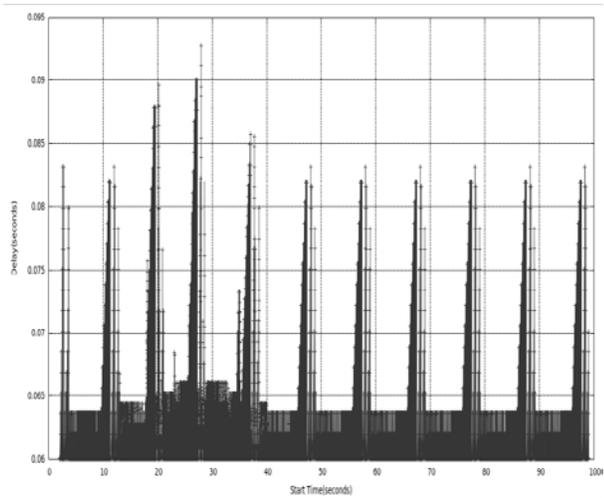

(b)SFQ

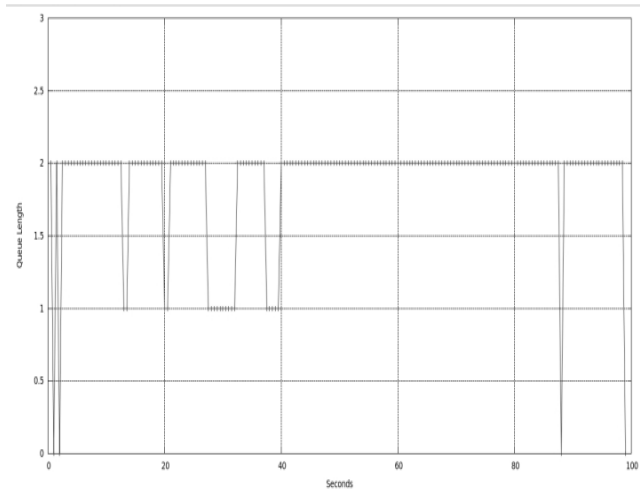

(b)SFQ

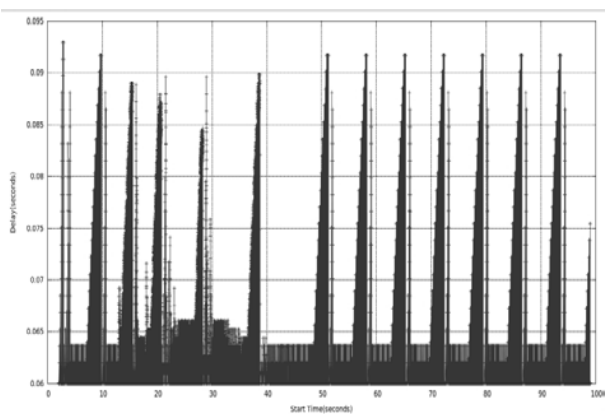

(c)REM

Fig. 6 Delay Diagram for various algorithms (a) RED, (b) SFQ, (c) REM

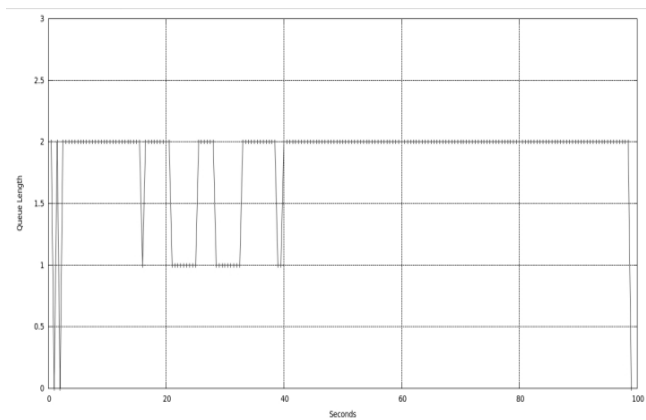

(c)REM

Fig. 7 Queue Length Diagram for various Algorithms (a) RED, (b) SFQ, (c) REM

Table 2  Comparative results

| Performance Metrics | | RED | SFQ | REM |
|---|---|---|---|---|
| Queue length | Max. | 2 | 2 | 2 |
| | Min. | 0 | 0 | 0 |
| Throughput | Max. | 5.53 | 6.64 | 7.51 |
| | Min. | 0 | 0 | 0 |
| Delay | Max. | 67.25 | 90.01 | 92.96 |
| | Min. | 60.03 | 60.03 | 60.03 |
| Send Packets | | 37157 | 42554 | 49117 |
| Lost Packets | | 151 | 56 | 66 |
| Average Loss Ratio (%) | | 0.4064 | 0.1316 | 0.1344 |
| Utilization (%) | | 59.45 | 68.08 | 78.58 |

Table 3 Ranking of the different algorithms

| Algorithm | Delay | Queue Length | Throughput | Loss Rate |
|---|---|---|---|---|
| RED | A | A | C | C |
| SFQ | B | B | B | A |
| REM | C | C | A | B |

3.4 Analysis of Delay

Figure 6 plots the actual response time for each packet achieved in RED, SFQ, and REM. It has been observed from Table 2 that minimum delay occurred in each algorithm is same but maximum delay achieved in REM. Therefore we could conclude that each algorithm would get a same response time provided congestion has been observed because queueing delay would be same for each algorithm if there is no congestion in network.

3.5 Analysis of Queue length

Here we did not achieve much difference in queue length between these algorithms because at most two packets could be allowed to enter into queue due to the small buffer capacity. REM achieved queue length of two packets for a longer time as shown in Figure 7.

## 4. Future Work And Conclusion

In this paper we address the problems with existing congestion control algorithms and we tried to show about various performance parameters of RED, SFQ, and REM for our considered network configurations. We have calculated the different performance parameters for each algorithm of considered network configuration as given in Figure 3 and Table 1. We calculated the total number of packets sent over the bottleneck link $R_1R_2$ and total number of packets lost during the simulation over the period of 100 seconds. SFQ has a minimum average loss ratio and RED has a maximum loss ratio. Now actual number of bytes transmitted over the bottleneck link $R_1R_2$ could be computed termed as utilization has been shown in Table 2. It has been observed that performance parameters are varying according to the algorithms. RED achieved the best result in terms of the delay but in terms of throughput, loss ratio, and utilization REM shows the best results. If we would provide the equal weightage to each performance parameter then we could conclude that REM would be the better one among all three algorithms considered in our simulation. Ranking for each performance parameter has been displayed in Table 3 as A indicates a higher ranking and ranking decrease up to C.

For future work, we plan to extend the simulation for the new algorithm which would comprise all the advantage of each algorithm. There would be hybridization of RED, SFQ, and REM to provide the better results.